\documentclass[twocolumn,secnumarabic,amssymb, nobibnotes, aps, prd]{revtex4-1}
\usepackage{xcolor}

\newcommand{\ie}[1]{{\it i.e.}}
\newcommand{\eg}[1]{{\it e.g.}}
\usepackage{graphicx}
 \usepackage{amsmath}
 \usepackage{array}[=2016-10-06]
 \usepackage{physics}
 
\begin{document}

\title{Generation of spin-squeezed states using dipole-coupled spins}%

\author{Yifan Song}%
\affiliation{Department of Chemistry, University of Southern California, Los Angeles, CA 90089, USA}

\author{Nabiha Hasan}%
\affiliation{Department of Chemistry, University of Southern California, Los Angeles, CA 90089, USA}

\author{Susumu Takahashi}%
\email[Corresponding email: ]{susumuta@usc.edu}
\affiliation{Department of Chemistry, University of Southern California, Los Angeles, CA 90089, USA}
\affiliation{Department of Physics \& Astronomy, University of Southern California, Los Angeles, CA 90089, USA}
\date{\today}%

%\tableofcontents
\begin{abstract}
Spins in solids and molecules are promising for applications of quantum sensing technology. 
    The sensitivity of the quantum sensing depends on how precisely spin observables can be determined in the measurement, and is intrinsically limited by the uncertainties of the observables.
    The use of a spin-squeezed state in a quantum sensor can reduce the uncertainty below the standard quantum limit when combined with an appropriate measurement procedure.
    Here, we discuss the simulation study of the generation of a squeezed state in an interacting spin system. 
    We show that a spin system coupled by the magnetic dipole interaction can create a squeezed state. Model systems to realize the spin squeezing experimentally are also discussed. In addition, we find that a squeezed state is a type of entangled state. 
    The present work paves the way to realize a squeezed state using a spin system to build a quantum sensor network with improved sensitivity, and to use it for the detection of quantum entanglement.
\end{abstract}
\maketitle

\section{Introduction}
Spins in solids and molecules have been investigated extensively for applications of quantum sensing techniques~\cite{degen2017quantum, aslam2023quantum, rovny2024nanoscale, crawford2021quantum, Yu-21}.
Spin-based quantum sensing senses the measurement quantity by encoding it into phase shifts of superposition or entangled spin states.
The quantum sensors have been shown to surpass the precision of classical sensors, offering improved accuracy and potential advantages such as miniaturization, which could lead to broader accessibility. These quantum sensing technologies have demonstrated the capability to measure a range of physical quantities with remarkable sensitivity, including magnetic~\cite{Degen2008, Maze2008, Balasubramanian2008, Taylor2008, Grinolds2013, Abeywardana2016, Fortman2020, Li2021, Ren2023} and electric fields~\cite{Dolde2011, Bian2021}, and temperature~\cite{Acosta2010, Toyli2013}.

The sensitivity of quantum sensing is ultimately limited by quantum fluctuations of the measurement observables where the fluctuations are rooted in the Heisenberg uncertainty principle.
For quantum sensing experiment based on an ensemble of spins, collective spin operators $J = (J_x, J_y, J_z)$ are the observables and governed by the cyclic commutation relations, \eg, $[J_y, J_z] = iJ_x$.
Therefore, the uncertainty relationship is given by $(\Delta J_y)_{\psi} (\Delta J_z)_{\psi} \geq 1/2 |\langle J_x \rangle_{\psi}|$ 
where $\langle ... \rangle_{\psi}$ is the expectation value with the state $\psi$ and $(\Delta J_y)_{\psi}$ is the uncertainty for $J_y$.
When $j$-axis ($k$-axis) is chosen by rotating $y$-axis ($z$-axis) along $x$-axis, and $(\Delta J_j)_{\psi} (\Delta J_k)_{\psi} = 1/2 |\langle J_x \rangle_{\psi}|$ with $(\Delta J_j)_{\psi}=(\Delta J_k)_{\psi}$ for any rotations, the value of $(\Delta J_j)_{\psi} (\Delta J_k)_{\psi}$ is minimum and the $(\Delta J_j)_{\psi}$ value is isotropic over the $(y,z)$ plane. 
In this situation, the measurement uncertainty is in the standard quantum limit (SQL).
Furthermore, the uncertainty value can be lowered. 
When $(\Delta J_j)_{\psi} (\Delta J_k)_{\psi} = 1/2 |\langle J_i \rangle_{\psi}|$ and $(\Delta J_j)_{\psi}>(\Delta J_k)_{\psi}$, the $(\Delta J_k)_{\psi}$ value is lower than that in SQL, and the corresponding state $\psi$ is called the spin squeezed state.
Quantum sensing can benefit from the squeezed state for sensitivity enhancement.
There are theoretical studies of spin squeezing,
including spin squeezing in a large ensemble of $S=1/2$ spins~\cite{Kitagawa1993, kuzmich1997spin, takeuchi2005spin, jin2009spin, Bennett2013, Huang_PRA_23}, and ferromagnetic systems~\cite{Block2024}, as well as several experimental demonstrations using spin-like systems~\cite{Takano2009, Hines2023, Borregaard_2017, Braverman2019, Qin2020}.
In addition, the implementation using spins in solids has been considered~\cite{Bennett2013}, however, it has yet to be explored experimentally.

In this paper, we discuss a method to create a spin-squeezed state from an interacting spin system with the magnetic dipole-dipole interaction.
The magnetic dipole-dipole interaction is a dominant interaction in dilute spins in solids and molecular systems. Because the interaction consists of a nonlinear Hamiltonian required to generate the squeezed state, and the strength of the interaction can be characterized quantitatively~\cite{Taminiau12, Kolkowitz12, Grinolds2014, Abeywardana2016, Li2021, Peng2019, Ren2024}, a dipolar-coupled spin system is potentially a great platform to realize a squeezed state. In this study, we first introduce the Hamiltonian of the dipole-coupled spin system and a method to calculate the uncertainty of spin operators of the system. In the present case, we consider a coherent spin state as an initial state. We then obtain the uncertainty of $J_y$ and $J_z$, \ie{}, $\Delta J_y$ and $\Delta J_z$. We consider cases with and without the magnetic dipole interaction, and show that a squeezed state emerges in the system with the dipole interaction. We also extend the discussion to a $N=2-10$ interacting spin system. The calculation result shows that the uncertainty is proportional to $1/\sqrt{N}$ for the system without the interaction, indicating that the uncertainty values are in SQL. Moreover, when the interaction is non-zero, the uncertainty bypasses SQL, which demonstrates the emergence of a squeezed state. Furthermore, we discuss triangle and linear three-spin systems as examples to realize the spin squeezed state.

\section{Interacting spin system}
\begin{figure}
    \centering
    \includegraphics[width=8 cm]{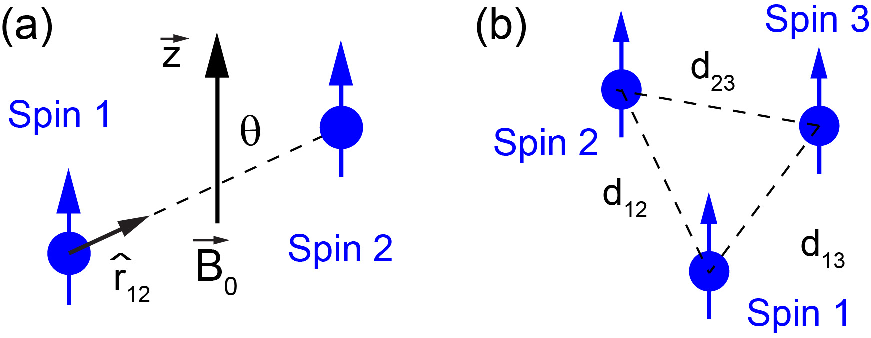}
    \caption{Models of interacting spin systems. (a) Dipole-coupled 2-spin system. (b) Dipole coupled 3-spin system.}
    \label{fig:n2n3}
\end{figure}
Here, we discuss the generation of squeezed states by considering an interacting spin system via a magnetic dipole-dipole interaction. 
First, we consider a pair of dipole-coupled spins, as shown in Fig.~\ref{fig:n2n3}(a) ($N$ = 2 where $N$ is the number of spins). The Hamiltonian of the system is given by,
\begin{equation}
H = H^1 + H^2 + H^{12},
\label{eq:totalH}
\end{equation}
where $H^i = \omega_0 S_z^i$ is the Zeeman term of the $i$-th spin ($i = 1$ or 2 in the present case). 
$\omega_0$ is the Zeeman energy to determine the Larmor frequency. 
In the present study, we consider a case where the Larmor frequencies of all spins are the same.
$S_z^i$ is the spin operator $S_z$ for the $i$-th spin. The third term $H^{12}$ is the Hamiltonian for the magnetic dipole interaction between Spin 1 and 2, namely,
\begin{equation}
H^{12} = (2\pi) \frac{\mu_0}{4\pi}  \frac{h^2 \gamma_1 \gamma_2}{r_{12}^3} [3 (S^1 \cdot \hat{r}_{12})(S^2 \cdot \hat{r}_{12})-S^1 \cdot S^2 ],    
\label{Eq:H_dip}
\end{equation}
where $\mu_0$ is the vacuum magnetic permeability. $h$ is the Planck constant. $\gamma_i$ is the gyromagnetic ratio for the i-th spin. For free electrons, $\gamma_i$ = 28 (GHz/Tesla). $S^i = (S_x^i, S_y^i, S_z^i)$.  $\hat{r}_{12}$ is a unit vector in the direction of the line joining the two spins as shown in Fig.~\ref{fig:n2n3}(a). 
In the present case, we consider that two spins have the same Larmor frequency, and the spins are aligned along the magnetic field (z-axis).
Then, using the secular approximation~\cite{Abragam1961, Mehring83}, Eq.~\ref{Eq:H_dip} can be rewritten as,
\begin{equation}
\begin{split}
H^{12} & = d_{12}  [S_z^1 S_z^2-1/4 (S_+^1 S_-^2+S_-^1 S_+^2 )] \\
& = d_{12} [S_z^1 S_z^2-1/2 (S_x^1 S_x^2+S_y^1 S_y^2 )],
\end{split}
\label{eq:H_dip2}
\end{equation}
where $d_{12}/(2\pi) =  \mu_0 / (4\pi) (\gamma_1 \gamma_2)/(r_{12}^3 ) |S|^2 (3 cos^2 \theta - 1)$. $\theta$ is the angle between the z-axis and $\hat{r}_{12}$ vector (see Fig.~\ref{fig:n2n3}(a)).
The dipole interaction strength $d_{12}$ depends on the distance $r_{12}$ and the angle $\theta$ between the spins. 
For instance, with $|r_{12}| = 1$ nm, $S = 1/2$, and $\theta = 0^\circ$, $d_{12}/(2\pi) = 26$ MHz is obtained.
Moreover, we transform the Hamiltonian in Eq.~\ref{eq:totalH} to a rotating frame at a frequency of $\omega$ using the unitary operator of $U = \rm{exp} (-i \omega t (S_z^1+S_z^2))$.
Since $U H^{12} U^\dag = H^{12}$ in the present case, the Hamiltonian in the rotating frame is given by, $H_r = (\omega_0 - \omega) J_z + H_r^{12}$ where $J_z = S_z^1 + S_z^2$. 
Therefore, with $\omega_0 - \omega = 0$, the rotating-frame Hamiltonian is given by,
\begin{equation}
H_r^{12} = d_{12} [S_z^1 S_z^2 - \frac{1}{2} (S_x^1 S_x^2+S_y^1 S_y^2 )].
\label{eq:H_12_r}
\end{equation}
Furthermore, for a $N$-spin system, the Hamiltonian can be extended to be $H = \Sigma_i [H^i+\Sigma_{j>i} H^{ij}]$ using Eq.~\ref{eq:H_12_r}.
Then, the rotating-frame Hamiltonian with $U = \exp(-i \omega_0 t \Sigma_i S_z^i)$ is given by, 
\begin{equation}
H_r  = \Sigma_{i=1}^N \Sigma_{j>i}^N d_{ij} [S_z^i S_z^j-1/2 (S_x^i S_x^j+S_y^i S_y^j )].
\label{eq:H_r}
\end{equation}
An example of $N = 3$ is shown in Fig.~\ref{fig:n2n3}(b). 
As can be seen, the $N = 3$ system is expressed with three dipole interaction strengths, \ie, $d_{12}$, $d_{13}$, and $d_{23}$.

Using Eq.~\ref{eq:H_r}, the time evolution of the state can be calculated easily. Since the Hamiltonian in Eq.~\ref{eq:H_r} is time-independent, using the Hamiltonian, the time evolution of the state ($|\psi(t)\rangle$) is given by,
\begin{equation}
|\psi(\tau)\rangle = \exp(-i H_r \tau) |\psi(0) \rangle,
\label{eq:state_tau}
\end{equation}
where $| \psi(0) \rangle$ is the initial state.
$\tau$ is the evolution time. 
As can be seen from Eq.~\ref{eq:state_tau}, we consider only coherent dynamics in the present study (spin decoherence processes are not considered).
Finally, using $|\psi(\tau)\rangle$ in Eq.~\ref{eq:state_tau}, the expectation value and the uncertainty of spin operators are given by,
\begin{equation}
\langle J_k \rangle = \langle\psi |J_k |\psi\rangle,
\label{eq:exp}
\end{equation}
and
\begin{equation}
\Delta J_k = \sqrt{(\langle \psi |(J_k )^2|\psi\rangle - \langle\psi |J_k |\psi\rangle^2},
\label{eq:uncer}
\end{equation}
respectively. $J_k$ ($k = x$, $y$, and $z$) is a spin operator for the total angular moment ($J_k = \Sigma_i S_k^i$). 
Moreover, by dividing $\Delta J_k$ in Eq.~\ref{eq:uncer} by the amplitude of the angular momentum ($J=\sqrt{( \langle J_x \rangle^2 + \langle J_y \rangle^2 + \langle J_z \rangle^2 }$), the normalized uncertainty of $\langle J_k \rangle$ is given by,
\begin{equation}
\sigma_k = \frac{\Delta J_k}{J},
\label{eq:SD}
\end{equation}
representing the inverse of the signal-to-noise ratio.
As described later, in the present study, $\langle J_y \rangle = \langle J_z \rangle = 0$ and $\langle J_x \rangle = J$. Therefore, using $\sigma_k$, the uncertainty relation is written by $\sigma_y \sigma_z \geq 1/(2 J)$. 
Moreover, for a non-entangled state where $J = N/2$, $\sigma_y \sigma_z \geq 1/N$. Thus, in SQL, $\sigma_y \sigma_z = 1/N$, and $\sigma_y = \sigma_z = 1/\sqrt{N}$.
In the present study, we use the normalized uncertainty value to determine the optimum spin squeezing conditions and the degree of squeezing.

Furthermore, for the discussion of spin-squeezed and entangled states, we introduce the von Neumann entropy for the $i$-th spin in a $N$ spin system~\cite{Nielsen_Chuang_2010}.
\begin{equation}
    S^i_{vN} = -Tr(\rho^i log(\rho^i)),
    \label{eq:entro}
\end{equation}
where $\rho^i$ is a partial trace of the density matrix ($\rho=\ket{\psi}\bra{\psi}$) for Spin-$i$ in the $N$ spin system. For example, the partial trace of Spin-1 in the $N=3$ system is given by $\rho^1 = Tr_{23}(\rho)$.
The partial trace is calculated by $\rho^A = Tr_{BC}(\ket{a_1}\bra{a_2}\otimes\ket{b_1 c_1}\bra{b_2 c_2}) = \ket{a_1}\bra{a_2} Tr(\ket{b_1 c_1}\bra{b_2 c_2})$~\cite{Nielsen_Chuang_2010}.
The von Neumann entropy is often used to probe quantum entanglement. 
When a quantum state is a superposition state, $S_{vN} = 0$. 
When a state is an entangled state, $S_{vN} > 0$, \eg, when the state is the Greenberger–Horne–Zeilinger (GHZ) state~\cite{GHZ-90},  $(\ket{\uparrow\uparrow\uparrow} + \ket{\downarrow\downarrow\downarrow}/\sqrt{2}$, the entropy value is $S^1_{vN} = S^2_{vN} = S^3_{vN} = log(2) = 0.69$.
We also note that we study coherent states only in the present investigation, and the effect of decoherence is not considered.

\section{Simulation procedure}
\begin{figure}
    \centering
    \includegraphics[width=8 cm]{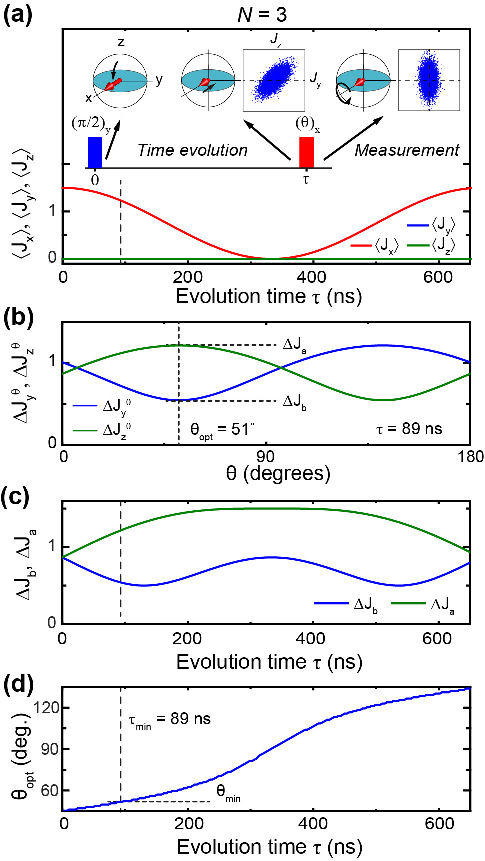}
    \caption{Calculation results of the expectation value, and uncertainty for $N = 3$ system. $d/(2\pi) = 1$ MHz. 
    (a) Expectation values ($\langle J_x \rangle$, $\langle J_y \rangle$, and $\langle J_z \rangle$) as a function of the evolution time ($\tau$). 
    The inset shows a pulse sequence to generate and measure the spin-squeezed state.
    (b) The uncertainty values as a function of angle $\theta$ at $\tau=89$ ns ($\Delta J_y^{\theta}$ and $\Delta J_z^{\theta}$). $\Delta J_b$ represents the minimum $\Delta J_y^{\theta}$. $\theta_{opt}$ is the corresponding angle for $\Delta J_b$. $\Delta J_a$ is the corresponding $\Delta J_z^{\theta}$ for $\theta_{opt}$.
    (c) $\Delta J_b^{\theta}$ and $\Delta J_a^{\theta}$ as a function of the evolution time.
    (d) $\theta_{opt}$ as a function of the evolution time. $\theta_{min}$ is the $\theta_{opt}$ value at $\tau_{min}$.
    The calculations were done with the step sizes of 1 ns and 1 degrees for $\tau$ and $\theta$, respectively.
    }
    \label{fig:exp_uncer}
\end{figure}
\begin{figure}
    \centering
    \includegraphics[width=8 cm]{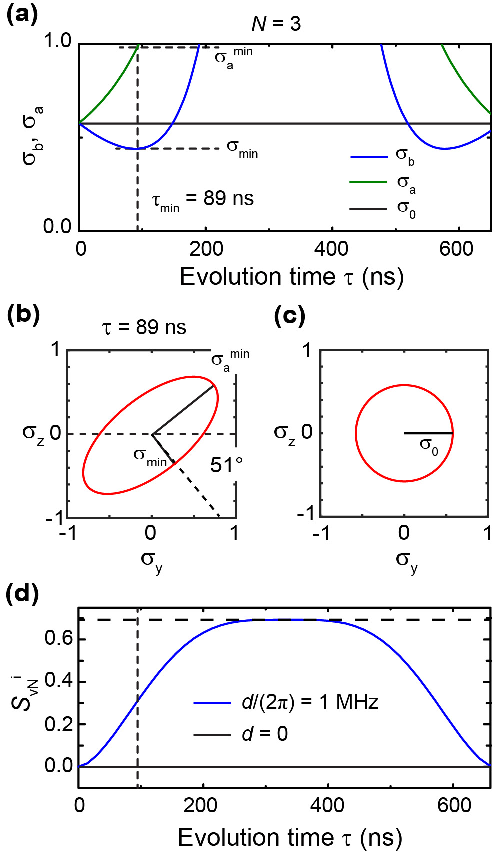}
    \caption{Emergence of spin-squeezed state. 
    (a) Normalized expectation values ($\sigma_b$ and $\sigma_a$) as a function of the evolution time ($\tau$). Black vertical dashed line indicates the optimum evolution time $\tau_{min}$ to give the minimum $\sigma_b$ ($\sigma_{min}$). $\sigma_{min} = 0.440$ at $\tau_{min} = 89$ ns.
    (b) Distribution of $\sigma_y$ and $\sigma_z$ for $d/(2\pi) = 1$ MHz. $\sigma_a^{min}=0.961$. (c) Distribution of $\sigma_y$ and $\sigma_z$ for $d = 0$. (d) $S^i_{vN}$ as a function of the evolution time $\tau$. $i=1-3$. $S^1_{vN}=S^2_{vN}=S^3_{vN}$ in the present case. Black vertical dashed line indicates the optimum evolution time. Black horizontal dashed line represents the $S^i_{vN}$ value of a fully entangled state.
    }
    \label{fig:norm_uncert}   
\end{figure}
In this section, we describe the simulation procedure to observe the spin squeezing. 
The inset of Fig.~\ref{fig:exp_uncer}(a) shows a pulse sequence, representing the procedure for the simulation and for experimental demonstration in the future.
As can be seen, we first prepare a coherent spin state using a $(\pi/2)$-pulse with the phase along the x-axis ($(\pi/2)_y$-pulse), therefore, the initial state is set by, $\ket{\psi(0)} = \otimes_{i=1}^N \ket{+x}_i=\otimes_{i=1}^N(\ket{0}_i + \ket{1}_i)/\sqrt{2}$, \ie, all spins are in the $|+x\rangle$ state in the Bloch sphere.
Then, we have the time evolution of the state with a duration of $\tau$.
In the simulation, the evolved state under the system Hamiltonian is calculated with Eq.~\ref{eq:H_r} and \ref{eq:state_tau}.
In the present simulation, we also consider the interaction strengths for all combinations to be the same in the calculation of this section, \ie, $d_{ij} = d$ in Eq.~\ref{eq:H_r}.
This condition is often used to study quantum dynamics in a weakly coupled system ~\cite{vidal2004entanglement}.
Next, we perform measurements with observables $J_y$ and $J_z$.
It is important to note that, as shown previously~\cite{Kitagawa1993}, when the state is in a squeezed state, the 2D distribution of the measured values with $J_y$ and $J_z$ becomes elliptical, and the semi-major and semi-minor axes of the ellipse are often tilted from the $J_y$ and $J_z$ axes.
Therefore, it is necessary to obtain the 2D map of the distribution.
In the present case, we do so by implementing a state rotation using a $(\theta)$-pulse along the $x$-axis at a time of $\tau$ ($(\theta)_x$-pulse), as shown in the inset of Fig.~\ref{fig:exp_uncer}(a).
The corresponding unitary operator of the rotation is $\ket{\psi(\tau)}_\theta = \exp(-i \theta J_x)\ket{\psi(\tau)}$.
After the state rotation, we perform the measurement.
In the simulation, we calculate the expectation values $\langle J_x \rangle$, $\langle J_y \rangle$, and $\langle J_z \rangle$, and the uncertainty values $\Delta J_y^{\theta}$ and $\Delta J_z^{\theta}$ using $\ket{\psi(\tau)}_\theta$ at each evolution time $\tau$ to obtain the mean values and the width of the distribution, respectively.
In addition, to find the semi-minor ($\Delta J_b$) and major ($\Delta J_a$) axes of the 2D map, $\Delta J_y^{\theta}$ and $\Delta J_z^{\theta}$ are calculated as a function of $\theta$. 
Then, as can be seen from Fig.~\ref{fig:exp_uncer}(b), $\Delta J_b$, $\Delta J_a$, and the corresponding angle ($\theta_{opt}$) are determined from the minimum value of $\Delta J_y^{\theta}$, and the corresponding values of $\Delta J_z^{\theta}$, and $\theta$, respectively.
Note that $\langle J_x \rangle$, $\langle J_y \rangle$, and $\langle J_z \rangle$ are independent of $\theta$.
Then, $\langle J_x \rangle$, $\langle J_y \rangle$, $\langle J_z \rangle$, $\Delta J_{b}$, $\Delta J_{a}$, and $\theta_{opt}$ are obtained as a function of the evolution time $\tau$ (see Fig.~\ref{fig:exp_uncer}(a),(c), and (d)).
Moreover, using $J$, $\Delta J_{b}$, $\Delta J_{a}$, and Eq.~\ref{eq:SD}, the normalized uncertainties ($\sigma_b$ and $\sigma_a$) are obtained as a function of $\tau$ (Fig.~\ref{fig:norm_uncert}(a)). 
By analyzing the $\sigma_{b}(\tau)$ value, we obtain the minimum normalized uncertainty $\sigma_{min}$ as well as the corresponding evolution time $\tau_{min}$, semi-major axis value $\sigma_{a}^{min}$ (see Fig.~\ref{fig:norm_uncert}(a)), and the optimum angle $\theta_{min}$ (Fig.~\ref{fig:exp_uncer}(d)).
We also note that the minimum uncertainty value for $d = 0$ ($\sigma_0$) is the same as $\sigma_{b}(\tau=0)$ and the SQL value.
In particular, the observation of $\sigma_{min}$ smaller than the SQL value of $1/\sqrt{N}$ represents the emergence of a spin-squeezed state.

\section{Simulation results}
Now, we discuss the simulation results. 
First, we consider a $N=3$ system with $d/(2\pi) = 1$ MHz. 
Fig.~\ref{fig:exp_uncer}(a)-(d) shows the calculation results of the expectation values, uncertainties, and optimum $\theta$ values. 
As can be seen, the expectation values for $J_y$ and $J_z$ are zero during the time evolution while the expectation value of $J_x$ varies from 1.5 to 0.
In particular, $\langle J_x \rangle$ becomes zero at $\sim$333 ns, indicating that the state is fully entangled.
Those expectation values are independent of $\theta$, as mentioned previously.
On the other hand, the uncertainty values depend on $\theta$ (see Fig.~\ref{fig:exp_uncer}(b)). 
For $\tau=89$ ns, the $\Delta J_y^{\theta}$ goes to the minimum value at $\theta = 51^\circ$.
This $\theta$ dependence on $\Delta J_y$ and $\Delta J_z$ indicates that the distribution shape of the uncertainty is elliptical, \ie{} emergence of the spin-squeezed state.
As shown in Fig.~\ref{fig:exp_uncer}(c), $\Delta J_b$ and $\Delta J_a$ vary as a function of the evolution time, indicating that the degree of the spin squeezing also varies.
In addition, we find that the optimum angle of $\theta$ varies with the evolution time, as shown in Fig.~\ref{fig:exp_uncer}(d).

Next, we discuss the normalized uncertainty ($\sigma_k$).
Figure~\ref{fig:norm_uncert}(a) shows the results for $N=3$. 
As can be seen, the normalized uncertainty shows that $\sigma_{b}$ takes the lowest value of 0.440 at the time evolution of $\tau_{min} = 89$ ns and $\theta_{min}$ of 51 degrees (see Fig.~\ref{fig:exp_uncer}(d)).
As can be seen, there are multiple minimum points of the normalized uncertainty with the same value, \eg{} the dip at $\tau=578$ ns, and we chose the one with the shortest evolution time in the present case.
As discussed in Sect.~\ref{append:intstr}, $\tau_{min}$ depends on the strength of the interaction. 
The larger the interaction strength, the smaller the $\tau_{min}$ value. 
In addition, the $\sigma_b$ and $\sigma_a$ values at $\tau=0$ are 0.577, which equals $1/\sqrt{3}$.
Namely, $\sigma_b$ $\sigma_a$ at $\tau = 0$ are in SQL. 
Therefore, at $\tau_{min}$, $\sigma_{min}$ is smaller than $\sigma_0$, showing the spin squeezing.
Figure~\ref{fig:norm_uncert}(a) also shows that the spin squeezing occurs in the ranges of $0-120$ and $540-650$ ns.
In addition, we show the 2D distribution of $\sigma_y$ and $\sigma_z$ at $\tau_{min} = 89$ ns in Fig.~\ref{fig:norm_uncert}(b).
As can be seen, the distribution is elliptical.
The semi-minor axis of the ellipse is $\sigma_{min}=0.440$ while the semi-major axis is $\sigma_a^{min}=0.961$.
The semi-minor axis of the ellipsoid is tilted from the $\sigma_y$ axis by an angle of $\theta_{min} = 51$ degrees.
Figure~\ref{fig:norm_uncert}(c) shows an isotropic distribution with $\sigma_0=0.577$ for $d = 0$, indicating the non-squeezed state.
Furthermore, we consider the nature of the spin-squeezed state by considering the relationship between the spin-squeezed state and an entangled state using the von Neumann entropy ($S_{vN}^i$) given in Eq.~\ref{eq:entro}. 
Figure~\ref{fig:norm_uncert}(d) shows $S_{vN}^i$ calculated for the $N=3$ system and $d/(2\pi) = 1$ MHz.
As can be seen, the entropy value starts from 0, and the entropy values increase during the time evolution, which shows the generation of quantum entanglement in the system. 
Around $\tau = 333$ ns, the entropy reaches the maximum value of 0.69, showing that the state is a fully entangled state.
This is consistent with $\langle J_x \rangle = 0$ at $\tau = 333$ ns.
By comparing with Fig.~\ref{fig:norm_uncert}(a) and (d), we can see that the squeezed state is a type of entangled state. 
However, the optimum squeezed state is not consistent with the maximally entangled state. 

\begin{figure}
    \centering
    \includegraphics[width=8 cm]{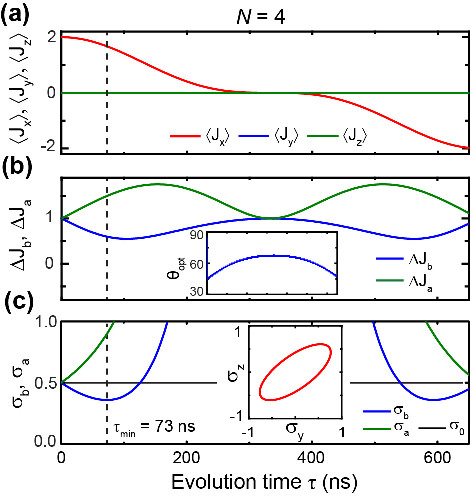}
    \caption{
    Spin squeezing of $N=4$ system with $d/(2\pi) = 1$ MHz. (a) Expectation values as a function of the evolution time $\tau$. Vertical black dashed line represents the evolution time of $\tau_{min} = 74$ ns. (b) $\Delta J_b$ and $\Delta J_a$ vs $\tau$. The inset shows $\theta_{opt}$ vs $\tau$. 
    (c) $\sigma_b$ and $\sigma_a$ as a function of $\tau$. Horizontal black solid line represents $\sigma_0$. The inset shows the 2D map of $\sigma_b$ and $\sigma_a$. $\sigma_{min} = 0.358$. $\sigma_a^{min} = 0.899$. $\tau_{min} = 73$ ns. $\theta_{min} = 54^{\circ}$.
    }
    \label{fig:n4}
\end{figure}
Moreover, figure~\ref{fig:n4} shows the results for $N=4$ and $d/(2\pi) = 1$ MHz.
As can be seen in Fig.~\ref{fig:n4}(a), $\langle J_x \rangle$ ranges from 2 to $-2$.
Also, around $\tau$ of 333 ns, $\langle J_x \rangle$ goes to zero, indicating that the states become fully entangled.
Figure~\ref{fig:n4}(b) shows that the uncertainty values ($\Delta J_{min}$ and $\Delta J_{max}$) oscillate as a function of the evolution time, and the optimum angle also varies (the inset of Fig.~\ref{fig:n4}(b)).
The normalized uncertainty value is also plotted in Fig.~\ref{fig:n4}(c). The minimum normalized uncertainty is given by $\sigma_{min}=0.358$ at $\tau_{min}=73$ ns.
As can be seen from Fig.~\ref{fig:ent} in Sect.~\ref{append:entropy}, the minimum uncertainty occurs with $S^i_{vN}$ of 0.290, showing that the corresponding squeezed state is a partially entangled state, but not a fully entangled state.
This uncertainty is smaller than that of SQL of 0.5. 
In addition, as shown in the inset of Fig.~\ref{fig:n4}(c), the distribution is elliptical, showing the emergence of the spin squeezing. 

\begin{figure}
    \centering
    \includegraphics[width=8 cm]{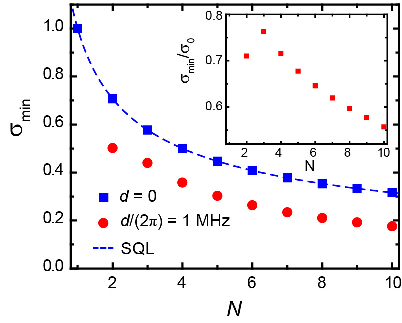}
    \caption{$\sigma_{min}$ as a function of the number of spin qubits ($N$) for cases with $d = 0$ and $d/(2\pi) = 1$ MHz. Blue dashed line indicates the standard quantum limit (SQL), \ie{} $\sigma_0= 1/\sqrt{N}$. The inset shows the ratio of $\sigma_{min}/\sigma_0$ vs $N$.}
    \label{fig:n3}    
\end{figure}

Furthermore, we performed the same set of calculations for $N = 2$ and $5 - 10$ systems. 
All calculation results are shown in Sect.~\ref{append:alln}.
The obtained amplitude of the expectation value ($J$), the normalized uncertainty ($\sigma_0$ and $\sigma_{min}$), the optimal angle ($\theta_{min}$) and the evolution time for $\sigma_{min}$ values ($\tau_{min}$) are also summarized in Table~\ref{tab:table1}.
Figure~\ref{fig:n3} plots $\sigma_{min}$ as a function of $N$.
As can be seen, the $\sigma_0$ values for $d = 0$ systems are consistent with a line representing SQL.
We also found that the $\sigma_{min}$ values of the interacting systems ($d/(2\pi) = 1$ MHz) are always lower than those of the non-interacting systems ($d = 0$), showing that the spin squeezing created by the magnetic dipole-dipole interaction can reduce the uncertainty in quantum measurement.
The ratio of the normalized uncertainty between the interacting and non-interacting systems ranges from 0.76 to 0.56.
The smaller $\sigma_{min}$ can be obtained from a system with larger $N$, as shown in the inset of Fig.~\ref{fig:n3}.
As can be seen from Fig.~\ref{fig:ent}, we also found that the minimum normalized uncertainty occurs with partially entangled states except for the case of $N=2$. 

\begin{table*}[t]
    \centering
    \begin{tabular}{||m{2.5cm} m{2.5cm} m{2.5cm} m{2.5cm} m{2.5cm} m{2.5cm}||}
    \hline
 $N$ & $J$ & $\sigma_0$ & $\sigma_{min}$ & $\tau_{min}$ (ns) & $\theta_{min}$ (degrees)\\ [0.8ex] 
 \hline\hline
 1 & 0.5 & 1.000 & - & - & - \\ 
 \hline
 2 & 1.0 & 0.707 & 0.500 & 333 & 45 \\ 
 \hline
 3 & 1.5 & 0.577 & 0.440 & 89 & 51 \\ 
 \hline
  4 & 2.0 & 0.500 & 0.358 & 73 & 54 \\ 
 \hline
  5 & 2.5 & 0.447 & 0.303 & 63 & 56 \\ 
 \hline
  6 & 3.0 & 0.408 & 0.263 & 57 & 58 \\ 
 \hline
  7 & 3.5 & 0.378 & 0.234 & 53 & 60 \\ 
 \hline
  8 & 4.0 & 0.353 & 0.210 & 49 & 61 \\ 
 \hline
  9 & 4.5 & 0.333 & 0.192 & 45 & 62 \\ 
 \hline
  10 & 5.0 & 0.316 & 0.176 & 43 & 63 \\ 
 \hline
 \end{tabular}
    \caption{Summary of the obtained results. $\sigma_{min}$: $J_y$ and $J_z$ rotation to give $\sigma_{min}$. $\tau_{opt}$: the evolution time to give $\sigma_{min}$}
    \label{tab:table1}
\end{table*}

\section{Interacting spin systems for experimental demonstration of spin-squeezed states.}
Here, we consider the following spin systems for the experimental demonstration to observe the spin squeezing from dipole-coupled spin systems.

\subsection{Calculation for dipole-coupled triangle system}\label{sect:trinagle}
\begin{figure}
    \centering
    \includegraphics[width=8 cm]{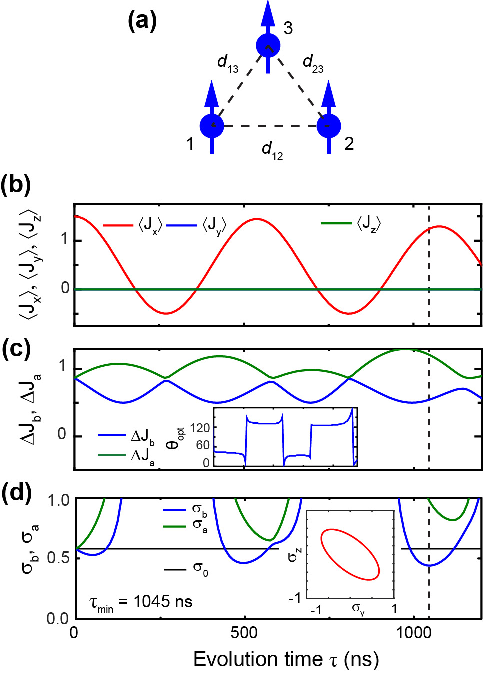}
    \caption{Triangle spins coupled via the magnetic dipole-dipole interaction. (a) Schematic of the triangle spin system. The coupling strengths are given by $d_{12}/(2\pi) = -1$ MHz, $d_{13}/(2\pi) = 1.6$ MHz, and $d_{23}/(2\pi) = 1.6$ MHz. (b) Expectation values of $Jx$, $J_y$, and $J_z$ as a function of evolution time. (c) $\Delta J_b$ and $\Delta J_a$ as a function of evolution time. The inset shows the corresponding angle of $J_b$. (d) The normalized uncertainty as a function of evolution time. $\sigma_{min}=0.440$. $\sigma^{min}_{a}=0.977$. $\tau_{min}=1045$ ns. $\theta_{min}=129$ degrees. The inset shows that the 2D distribution of $\sigma_b$ and $\sigma_a$.}
    \label{fig:triangle}
\end{figure}
The first system is a dipole-coupled triangle spin system (see Fig.~\ref{fig:triangle}(a)).
We consider that all spins are directed in the same direction.
A desired triangle spin system may be prepared as a molecular spin because of a flexible bottom-up approach to molecular synthesis.
A ferromagnetically coupled triangle molecular spin has also been investigated using pulsed EPR spectroscopy ~\cite{Abeywardana20162}.
As shown in Fig.~\ref{fig:triangle}(a), we consider that the system forms a right triangle, and all spins are directed along the $+z$-direction.
Using Eq.~\ref{eq:H_r}, the rotating frame Hamiltonian of the system is given by, $H_r = H_r^{12} + H_r^{13} + H_r^{23}$, where $H_r^{ij}$ is defined in Eq.~\ref{eq:H_12_r}.
By setting $d_{12}/(2\pi) = -1$ MHz, both $d_{13}/(2\pi)$ and $d_{23}/(2\pi)$ are given to be 1.6 MHz (see Fig.~\ref{fig:triangle}(a)). 
Using the same procedure as described in the previous section, we calculated the expectation values, uncertainty values, and the normalized uncertainty values. 
The results are summarized in Fig.~\ref{fig:triangle}(b). 
Figure~\ref{fig:triangle}(b) shows the time evolution of the expectation values. The expectation values of $J_x$ vary from 1.5 to -0.5. In addition, as can be seen, $\langle J_x \rangle$ goes to zero, showing that the states become fully entangled. 
Figure~\ref{fig:triangle}(c) shows that $\Delta J_b$ and $\Delta J_a$ values are different and change over the time evolution,  indicating that the state becomes squeezed.
$\sigma_b$ and $\sigma_a$ as a function of the evolution time are plotted in Fig.~\ref{fig:triangle}(c).
As can be seen, the $\sigma_b$ values are smaller than $\sigma_0$ around 50, 500 and 1050 ns, and $\sigma_b$ gives the minimum $\sigma_b$ value ($\sigma_{min}$) of 0.440 at $\tau = 1045$ ns.
Using those the $\sigma_{min}$ and $\sigma_a$ values, the 2D distribution of $\sigma_b$ and $\sigma_a$ is shown in the set of Fig.~\ref{fig:triangle}(c).
As shown, the distribution shape is elliptical, showing the emergence of the squeezed state. 
$\sigma_{min}$ of 0.440 give 24\% smaller than that of SQL for $N=3$.

\subsection{Calculation for dipole-coupled linear 3-spin system}
\begin{figure}
    \centering
    \includegraphics[width=8 cm]{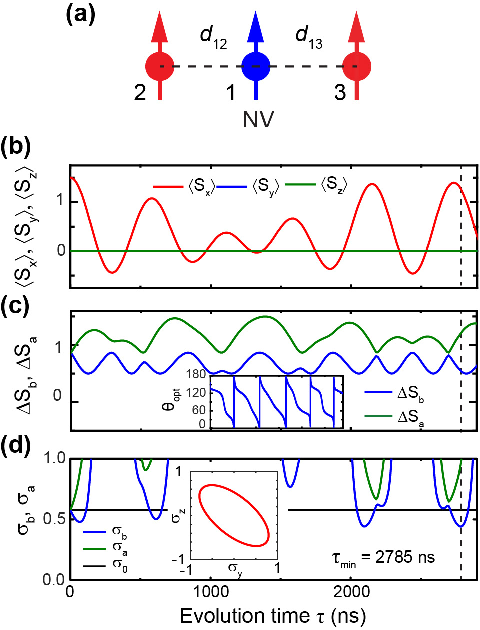}
    \caption{Three spin chain system. (a) Diagram of a three-spin chain based on a single NV center and two surrounding spins. 3-spin chain system with the coupling strengths of $d_{12}/(2\pi) = -1$ MHz, $d_{13}/(2\pi) = -2$ MHz. (b) Simulated expectation values of $S_x$, $S_y$, and $S_z$ as a function of evolution time. (c) $\Delta S_b$ and $\Delta S_a$ as a function of evolution time. The inset shows the corresponding angle for $\Delta J_b$. (d) The distribution of the normalized uncertainty as a function of evolution time. The inset shows the 2D distribution of $\sigma_b$ and $\sigma_a$. $\sigma_{min} = 0.440$. $\sigma_a^{min} = 0.959$. $\tau_{min} = 2785$ ns. $\theta_{min} = 129$ degrees.
    }
    \label{fig:chain}
\end{figure}
The second candidate system is a spin chain system as shown in Fig.~\ref{fig:chain}(a).
Similar to the system discussed in Sect.~\ref{sect:trinagle}, we consider that all spins are directed in the same direction. 
The considered spin chain consists of three spins, including a central electron spin coupled to two neighboring electron spins.
The system models a single NV center coupled to two neighboring electron spins.
Such a NV spin system has been reported recently~\cite{Ren2024}. 
The rotating frame Hamiltonian is given by, $H_r = H_r^{12} + H_r^{13}$, where $H_r^{ij}$ is defined in Eq.~\ref{eq:H_12_r}. 
In the calculation, we used $d_{12}/(2\pi) = -1$ MHz and $d_{13}/(2\pi) = -2$ MHz, which is similar to those from the recent study~\cite{Ren2024}. 
We also considered that the Larmor frequencies of the NV center and the two spins are the same. 
This condition can be achieved by setting the proper strength and orientation of a magnetic field, \ie{} a magnetic field at approximately 50 mT along the [111]-direction of the diamond lattice~\cite{Epstein2005}.
In addition, since the observables of the system are NV’s spin operators ($S_x^1$, $S_y^1$ and $S_z^1$ where the superscription "1" indicates the index of the spin shown in Fig.~\ref{fig:chain}(a)), we calculated the normalized uncertainty of $S_y^1$ and $S_z^1$.
Figure~\ref{fig:chain}(b) shows the expectation values. 
Similar to the previous case, the $S_x$ varies over the evolution time, $S_y$ and $S_z$ stay zero.
Simulation results for $\Delta S_b$ and $\Delta S_a$ are plotted in Fig.~\ref{fig:chain}(c). Their corresponding $\theta_{opt}$ angles are shown in the inset of Fig.~\ref{fig:chain}(c).
The calculated normalized uncertainty values are shown in Fig.~\ref{fig:chain}(d). 
As can be seen, $\sigma_b$ is smaller than $\sigma_0$ around 70, 600, 2100, and 2800 ns. $\sigma_b$ also becomes the minimum value of $\sigma_{min} = 0.440$ at $\tau=2785$ ns.
In addition, as plotted in the inset of Fig.~\ref{fig:chain}(d), the distribution shape is elliptical, showing the emergence of the squeezed state. The squeezing reduces the normalized uncertainty by 24$\%$ from that in SQL for the $N=3$ system.

\section{Summary}
In summary, we discussed the spin squeezing generated by spin systems interacting with the magnetic dipole-dipole interactions. Using the spin-squeezed states, we showed that the normalized uncertainty can be reduced to a value smaller than the standard quantum limit. 
We also analyzed the nature of the spin-squeezed state with the von Neumann entropy. The analysis shows that the spin-squeezed state is closely related to the entangled state. 
While the spin-squeezed state is always an entangled state, the degree of the squeezing is inconsistent with the degree of the entanglement. 
This finding suggests that spin squeezing can serve as a diagnostic probe to find the entanglement, providing a new tool for exploring quantum entanglement and correlations in ensemble spins. 
We further discussed two spin systems, triangle and linear three-spin systems, which may be used for experimental demonstration of the spin-squeezed state. 
The linear system can be realized with a single NV center coupled to two electron spins as the NV center coupled to two electron spins has been reported recently~\cite{Ren2024}.
The present approach, based on dipole-coupled spin systems, sheds light on demonstrating the spin-squeezed state experimentally and utilizing it for applications of quantum sensing and other quantum methodologies.
Moreover, the efficiency of spin squeezing generation is potentially enhanced by integrating our approach with time-dependent controls, such as optimized pulse sequences~\cite{khaneja2005optimal, khaneja2001time,Duan-13, li2011optimal}, and machine learning-assisted protocols~\cite{Koutromanos-24, duan2025concurrent,chen2019extreme,tan2021generation}, and by testing them using a versatile quantum device such as the IBM quantum computer~\cite{garcia2020ibm}.
Furthermore, the present study can be extended to take into account the effects of decoherence and population relaxations for the design of spin-squeezed states for applications of quantum sensing techniques in the future.
During the preparation of the manuscript, we noticed a recent preprint by Wu {\it et al.}, reporting spin squeezing in an ensemble of NV centers in diamond~\cite{wu2025spin}.

\section*{Acknowledgments}
This work was supported by the National Science Foundation (Grant Nos. CHE-2404463, and CHE-2004252 with partial co-funding from the Quantum Information Science program in the Division of Physics), and the Zumberge Foundation (ST).

\section*{Data Availability}
The data that support the findings of this article are not publicly available. The data are available from the authors upon reasonable request.

\appendix
\section{Dependence of the interaction strength}\label{append:intstr}
We show the time evolution of the minimum normalized uncertainty ($\sigma_{min}$) for the $N = 3$ system with $d/(2\pi) = 0.1$, 1, and 10 MHz. As shown in Fig.8, the system with a larger interaction strength can reach the minimum $\sigma_b$ faster, while the minimum values are the same for the three cases.
\begin{figure}
    \centering
    \includegraphics[width=8 cm]{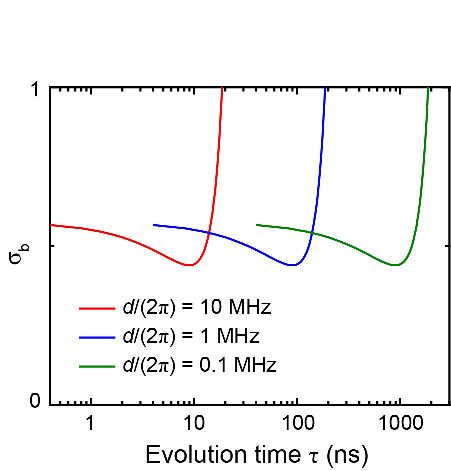}
    \caption{$\sigma_{b}$ as a function of the evolution time $\tau$ for systems with $d/(2\pi) = 0.1$, 1, and 10 MHz.}
    \label{fig:IntStr}
\end{figure}

\section{von Neumann entropy}~\label{append:entropy}
\begin{figure}
    \centering
    \includegraphics[width=8 cm]{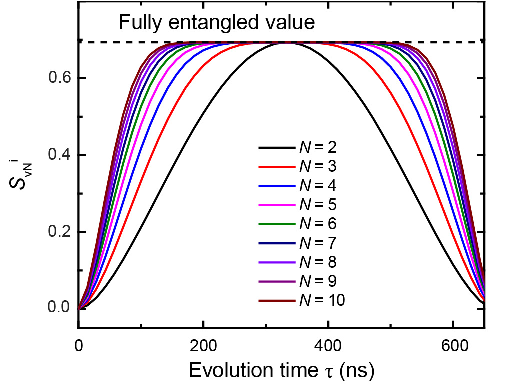}
    \caption{The entropy $S^i_{vN}$ as a function of evolution time $\tau$ for $N=2$ to 10. $i = 1$ to $N$. In the present case, the obtained $S^i_{vN}$ values are independent of $i$.}
    \label{fig:ent}
\end{figure}
The calculation of the von Neumann entropy ($S_{vN}^i$) for $N=2-10$ spin systems vs the evolution time $\tau$ is shown in Fig.~\ref{fig:ent}. In the calculation, all spins are coupled by the dipole interaction written in Eq.~\ref{Eq:H_dip} and the strengths of the interactions are $d/(2\pi) = 1$ MHz.
The entropy values for all sites (from $S_{vN}^1 (\tau)$ to $S_{vN}^N (\tau)$) are the same.
As shown, all cases reach the value of 0.69, indicating all cases reach fully entangled states. 
Moreover, the larger the number of spins, the faster it reaches the maximum $S_{vN}^i (\tau)$. 
For larger $N$, $S_{vN}^i (\tau)$ stays at the maximum for a longer duration.

\section{Results for $N=2$, and $N=5-10$}~\label{append:alln}
Here we summarize the results for $N=2$ and $N=5-10$ systems. The expectation values, the uncertainty, the normalized uncertainty, and the optimum angles as a function of the evolution time are presented. The 2D distributions of $\sigma_y$ and $\sigma_z$ are also shown.
As described earlier, these values are calculated using Eqs.~\ref{eq:exp}, \ref{eq:uncer}, and \ref{eq:SD}.
In the calculation, the initial state is $\ket{\psi(0)} = \otimes_{i=1}^N \ket{+x}_i=\otimes_{i=1}^N(\ket{0}_i + \ket{1}_i)/\sqrt{2}$.
The strengths of the dipolar interactions are considered to be the same, and $d/(2\pi) = 1$ MHz.

\begin{figure}
    \centering
    \includegraphics[width=8 cm]{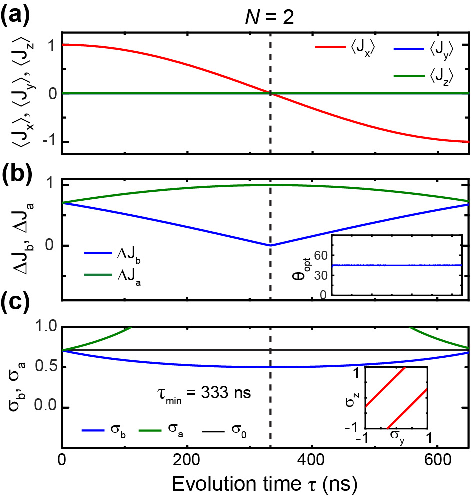}
    \caption{
    Spin squeezing of $N=2$ system with $d/(2\pi) = 1$ MHz. 
    (a) Expectation values as a function of the evolution time $\tau$. Vertical black dashed line represents the evolution time of $\tau_{min} = 333$ ns. (b) $\Delta J_b$ and $\Delta J_a$ vs $\tau$. The inset shows $\theta_{opt}$ vs $\tau$. 
    (c) $\sigma_b$ and $\sigma_a$ as a function of $\tau$. Horizontal black solid line represents $\sigma_0$. The inset shows the 2D map of $\sigma_b$ and $\sigma_a$. $\sigma_{min} = 0.500$. $\sigma_a^{min} = 637$. $\tau_{min} = 333$ ns. $\theta=45^{\circ}$. 
    }
    \label{fig:n2}
\end{figure}

The results for $N=2$ are presented in Fig.~\ref{fig:n2}.
As shown in Fig.~\ref{fig:n2}(a), $\langle J_x \rangle$ ranges from 1 to $-1$ while $\langle J_y \rangle = \langle J_z \rangle =0$ over the evolution time of $0 - 650$ ns.
At $\sim 333$ ns, the $\langle J_x \rangle$ value goes to zero, indicating that the state is fully entangled.
Moreover, as shown in Fig.~\ref{fig:n2}(b), the uncertainty values decrease from $\tau=0$ to 333 ns, and also become zero at $\sim$333 ns.
In addition, the optimum angle stays at 45 degrees over the evolution time as shown in the inset of Fig.~\ref{fig:n2}(b).
Figure~\ref{fig:n2}(c) shows the result of the normalized uncertainty.
As can be seen, the $\sigma_b$ value is lower than that of SQL ($\sigma_0$) in most ranges of the time evolution. 
Based on our criteria (the minimum $\sigma_b$), we found that $\sigma_{min} = 0.500$ at $\tau$ of 333 ns.
We also noticed that the state at $\tau = 333$ ns is fully entangled, which may not be practical for quantum measurements. 
In an experiment, it may be more practical to use the squeezed states in the ranges of $0-200$ and $450-650$ ns where the $\langle J_x \rangle$ values are non-zero.

\begin{figure}
    \centering
    \includegraphics[width=8 cm]{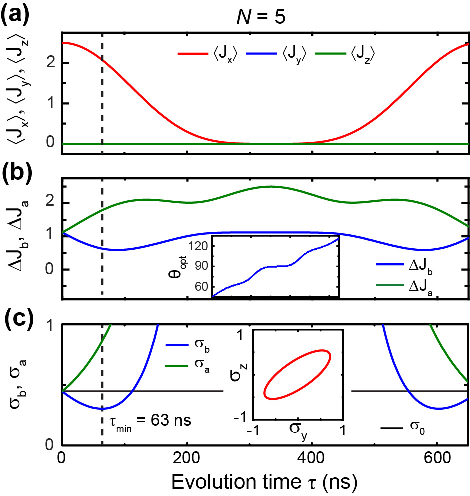}
    \caption{
    Spin squeezing of $N=5$ system with $d/(2\pi) = 1$ MHz. 
    (a) Expectation values as a function of the evolution time $\tau$. Vertical black dashed line represents the evolution time of $\tau_{min} = 63$ ns. 
    (b) $\Delta J_b$ and $\Delta J_a$ vs $\tau$. The inset shows $\theta_{opt}$ vs $\tau$. 
    (c) $\sigma_b$ and $\sigma_a$ as a function of $\tau$. Horizontal black solid line represents $\sigma_0$. The inset shows the 2D map of $\sigma_b$ and $\sigma_a$. $\sigma_{min} = 0.303$. $\sigma_a^{min} = 0.852$. $\tau_{min} = 63$ ns. $\theta=56^{\circ}$.
    }
    \label{fig:n5}
\end{figure}

Figure~\ref{fig:n5} shows the results for the $N=5$ spin system.
Figure~\ref{fig:n5}(a) shows the expectation values as a function of the evolution time. As shown, $\langle J_x \rangle$ ranges from 2.5 to 0 while $\langle J_y \rangle = \langle J_z \rangle =0$.
Similar to $N=2-4$, $\langle J_x \rangle$ also becomes zero around 333 ns, indicating that states are fully entangled.
As shown in Fig.~\ref{fig:n5}(b), the uncertainty values vary over the time evolution. The optimum angle increases from 45 to 90 degrees in the range from 0 to 300 ns, as shown in the inset of Fig.~\ref{fig:n5}(b).
Figure~\ref{fig:n5}(c) shows $\sigma_b$ and $\sigma_a$.
As can be seen, $\sigma_b$ becomes lower than that of SQL ($\sigma$) in the ranges of $0-100$ and $600-650$ ns.
We found that the optimum evolution time is 63 ns, giving $\sigma_{min}=0.303$.
The inset of Fig.~\ref{fig:n5}(c) shows that an elliptical distribution of the uncertainty, indicating the emergence of the spin-squeezed state.

\begin{figure}
    \centering
    \includegraphics[width=8 cm]{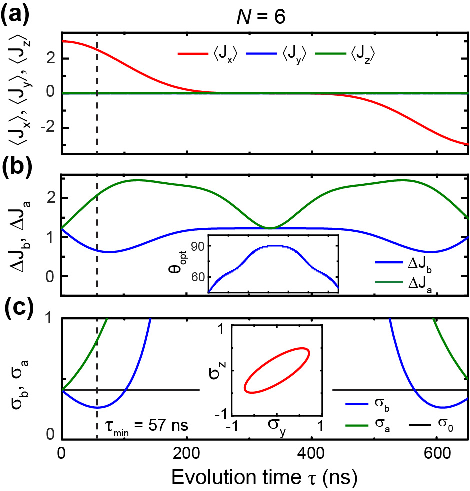}
    \caption{
    Spin squeezing of $N=6$ system with $d/(2\pi) = 1$ MHz. 
    (a) Expectation values as a function of the evolution time $\tau$. Vertical black dashed line represents the evolution time of $\tau_{min} = 57$ ns. 
    (b) $\Delta J_b$ and $\Delta J_a$ vs $\tau$. The inset shows $\theta_{opt}$ vs $\tau$. 
    (c) $\sigma_b$ and $\sigma_a$ as a function of $\tau$. Horizontal black solid line represents $\sigma_0$. The inset shows the 2D map of $\sigma_b$ and $\sigma_a$. $\sigma_{min} = 0.263$. $\sigma_a^{min} = 0.827$. $\tau_{min} = 57$ ns. $\theta=58^{\circ}$.
    }
    \label{fig:n6}
\end{figure}

\begin{figure}
    \centering
    \includegraphics[width=8 cm]{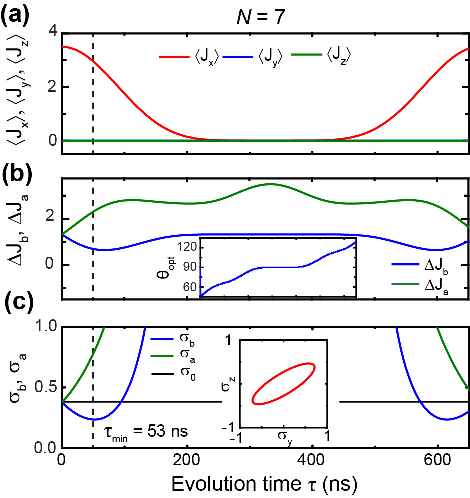}
    \caption{
    Spin squeezing of $N=7$ system with $d/(2\pi) = 1$ MHz. 
    (a) Expectation values as a function of the evolution time $\tau$. Vertical black dashed line represents the evolution time of $\tau_{min} = 53$ ns. 
    (b) $\Delta J_b$ and $\Delta J_a$ vs $\tau$. The inset shows $\theta_{opt}$ vs $\tau$. 
    (c) $\sigma_b$ and $\sigma_a$ as a function of $\tau$. Horizontal black solid line represents $\sigma_0$. The inset shows the 2D map of $\sigma_b$ and $\sigma_a$. $\sigma_{min} = 0.234$. $\sigma_a^{min} = 0.814$. $\tau_{min} = 53$ ns. $\theta=60^{\circ}$.
    }
    \label{fig:n7}
\end{figure}

\begin{figure}
    \centering
    \includegraphics[width=8 cm]{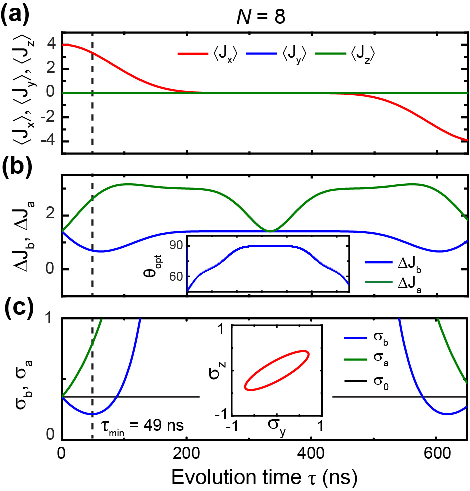}
    \caption{
    Spin squeezing of $N=8$ system with $d/(2\pi) = 1$ MHz. 
    (a) Expectation values as a function of the evolution time $\tau$. Vertical black dashed line represents the evolution time of $\tau_{min} = 49$ ns. 
    (b) $\Delta J_b$ and $\Delta J_a$ vs $\tau$. The inset shows $\theta_{opt}$ vs $\tau$. 
    (c) $\sigma_b$ and $\sigma_a$ as a function of $\tau$. Horizontal black solid line represents $\sigma_0$. The inset shows the 2D map of $\sigma_b$ and $\sigma_a$. $\sigma_{min} = 0.210$. $\sigma_a^{min} = 0.793$. $\tau_{min} = 49$ ns. $\theta=61^{\circ}$.
    }
    \label{fig:n8}
\end{figure}

\begin{figure}
    \centering
    \includegraphics[width=8 cm]{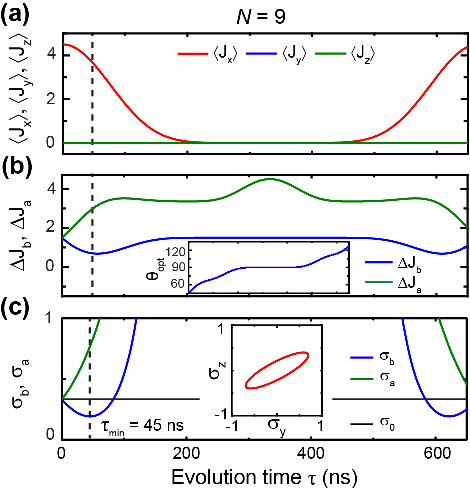}
    \caption{
    Spin squeezing of $N=9$ system with $d/(2\pi) = 1$ MHz. 
    (a) Expectation values as a function of the evolution time $\tau$. Vertical black dashed line represents the evolution time of $\tau_{min} = 45$ ns. 
    (b) $\Delta J_b$ and $\Delta J_a$ vs $\tau$. The inset shows $\theta_{opt}$ vs $\tau$. 
    (c) $\sigma_b$ and $\sigma_a$ as a function of $\tau$. Horizontal black solid line represents $\sigma_0$. The inset shows the 2D map of $\sigma_b$ and $\sigma_a$. $\sigma_{min} = 0.192$. $\sigma_a^{min} = 0.767$. $\tau_{min} = 45$ ns. $\theta=62^{\circ}$.
    }
    \label{fig:n9}
\end{figure}

\begin{figure}
    \centering
    \includegraphics[width=8 cm]{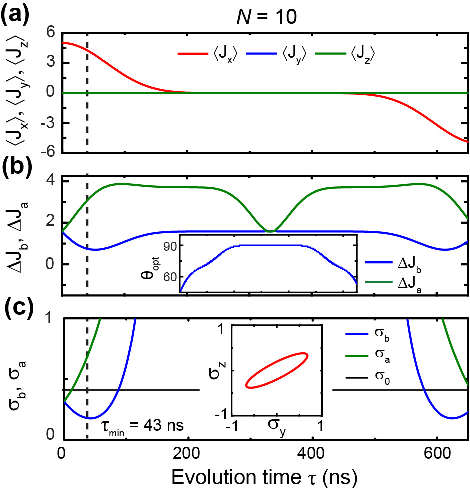}
    \caption{
    Spin squeezing of $N=10$ system with $d/(2\pi) = 1$ MHz. 
    (a) Expectation values as a function of the evolution time $\tau$. Vertical black dashed line represents the evolution time of $\tau_{min} = 43$ ns. 
    (b) $\Delta J_b$ and $\Delta J_a$ vs $\tau$. The inset shows $\theta_{opt}$ vs $\tau$. 
    (c) $\sigma_b$ and $\sigma_a$ as a function of $\tau$. Horizontal black solid line represents $\sigma_0$. The inset shows the 2D map of $\sigma_b$ and $\sigma_a$. $\sigma_{min} = 0.176$. $\sigma_a^{min} = 0.761$.  $\tau_{min} = 43$ ns. $\theta=63^{\circ}$.
    }
    \label{fig:n10}
\end{figure}

For $N=6-10$ spin systems, the results are presented in Fig.~\ref{fig:n6} to \ref{fig:n10}. 
As commonly seen in all cases, $\langle J_x \rangle$ also becomes zero around 333 ns, indicating that states are fully entangled.
In addition, those cases show that their $\sigma_{min}$ values are smaller than those of SQL.
Therefore, all cases show the emergence of spin squeezing.
Furthermore, we noticed that a systematic difference between odd and even numbers of spins, as can be seen in Fig~\ref{fig:exp_uncer}(a), Fig.~\ref{fig:n4}(a), and Sect.~\ref{append:entropy}. 
For example, the expectation values for even spins are positive and negative values, while the expectation values for odd spins are always positive.
The evolution time dependence on $\theta_{opt}$ for even spins is also symmetric, while that for odd spins is not.
While the origin of the difference is out of the scope in the present study, an odd-even dependence in a finite-size spin system has been reported previously.~\cite{Haas-2000}

\newpage
%\bibliography{ref}


\end{document}